# Deeply decarbonizing residential and urban central districts through photovoltaics plus electric vehicle applications[1]

Takuro Kobashi[1]*, Younghun Choi[1], Yujiro Hirano[2], Yoshiki Yamagata[1], Kelvin Say[3]

[1]Center for Global Environmental Research, National Institute for Environmental Studies, 16-2 Onogawa, Tsukuba, Ibaraki, 305-8506 Japan.

[2]Center for Social and Environmental Systems Research, National Institute for Environmental Studies, 16-2 Onogawa, Tsukuba, Ibaraki, 305-8506 Japan.

[3]Energy Transition Hub, University of Melbourne, Parkville, 3010, Australia.

*Corresponding author: kobashi.takuro@nies.go.jp

**Abstract**

With the costs of renewable energy technologies declining, new forms of urban energy systems are emerging that can be established in a cost-effective way. The "SolarEV City concept" has been proposed that uses rooftop Photovoltaics (PV) to its maximum extent, combined with Electric Vehicle (EV) with bi-directional charging for energy storage. Urban environments consist of various areas, such as residential and commercial districts, with different energy consumption patterns, building structures, and car parks. The cost effectiveness and decarbonization potentials of "PV + EV" and "PV (+ battery)" systems vary across these different urban environments and change over time as cost structures gradually shift. To evaluate these characteristics, we performed techno-economic analyses of PV, battery, and EV technologies for a residential area in Shinchi, Fukushima and the central commercial district of Kyoto, Japan between 2020 and 2040. We found that "PV + EV" and "PV only" systems in 2020 are already cost competitive relative to existing energy systems (grid electricity and gasoline car). In particular, the "PV + EV" system rapidly increases its economic advantage over time, particularly in the residential district which has larger PV capacity and EV battery storage relative to the size of energy demand. Electricity exchanges between neighbors (e.g., peer-to-peer or microgrid) further enhanced the economic value (net present value) and decarbonization potential of "PV + EV" systems up to 23% and 7% in 2030, respectively. These outcomes have important strategic implications for urban decarbonization over the coming decades.

---

[1] The short version of the paper was presented at ICAE2020, Dec 1-10, 2020. This paper is a substantial extension of the short version of the conference paper.

**Nomenclature**

| Abbreviations | | NPV | Net Present Value |
|---|---|---|---|
| 3D | 3 Dimensions | PV | Photovoltaics |
| CS | Cost Saving | SAM | System Advisor Model |
| ES | Energy Sufficiency | SC | Self-Consumption |
| EV | Electric Vehicles | SoC | State of Charge |
| FITs | Feed-in-tariffs | SPB | Simple Payback Period |
| GIS | geographic information system | SS | Self-Sufficiency |
| HVAC | Heating and ventilation, and air conditioning | V2C | Vehicle to Community |
| ICE | Internal Combustion Engine | V2H | Vehicle to Home |
| IRR | Internal Return Rate | VRE | Variable Renewable Energy |

## 1. Introduction

### 1.1. Cost effective urban decarbonization with renewable energy

Effective pathways to decarbonize urban energy system are constantly changing with new technologies and declining costs, occurring at different rates in different regions owing to climate, tariff structures, and regulations [1–3]. Over the next 10-20 years, it is expected that solar photovoltaics (PV) and battery costs will continue to decline [4,5]. As a result, the number of Electric Vehicles (EVs) are likely to increase substantially in urban environments [4,6]. This outcome creates an unprecedented opportunity to cost-effectively build near-zero-emissions power systems within our cities by coupling PV with EV as energy storage, which is named as the "SolarEV City concept" [7,8].

In Japan, Vehicle-to-Home (V2H) systems (i.e., bi-directional charging for the home and EV) was first commercialized in 2012 by Nichicon in a wake of the 2011 Fukushima Daiichi nuclear disaster that led to the increased needs for decentralized power supply. As of 2020, 8,500 V2H units have been sold (of which 1,800 units were sold in 2020 alone). This corresponds to 18% of new Nissan LEAF sales in 2020 [9]. Although the absolute number is still small, it is expected that the number of V2H sale will rapidly increase in line with the growth in EV penetration in Japan over the coming years (partly supported by subsidies). The drivers for V2H adoption are also related to changes in Feed-In Tariffs (FITs) for household roof-top PV system. Since 2019 (and after 10 years since their first introduction in 2009) FITs for residential PV systems have started to expire. To improve the value of excess PV electricity, car owners have begun switching to EV with a V2H system to be used as energy storage and a backup power supply [9].

With PV-V2H systems being recognized as a cost-effective option to decarbonize urban energy systems, while also providing an additional source of electricity in times of disaster [10], the government of Japan introduced in 2021 capital subsidies for V2H (max $6,800 for purchase + max

$3,600 for installation) and EVs (max $7,300 for purchase) in addition to tax credits. Notably, local governments have also provided further subsidies for V2H and EVs in Japan. With the government declaration of electrification of all new cars by 2035, Japan is likely going to experience a rapid increase of EVs, which is the basis for building the SolarEV City concept.

In our earlier studies, we found that households in Kyoto can benefit from having PV systems integrated with EVs as battery with substantial cost savings and $CO_2$ emission reduction from 2030 onwards [2]. As a whole city, Kyoto can save 22–37% of energy costs by deploying PV on up to 70% of rooftops across Kyoto City and using EVs as storage. This results in $CO_2$ emissions reducing 60–74% from greater renewable energy usage and removal of gasoline combustion [8]. However, it was not clear how and where building such a system should be initiated and expanded across an urban environment, which is an important strategic consideration when planning for a net-zero emission society by 2050.

*1.2. Building and district demand analysis with 3D energy balance model*

Cities are not homogenous but have a wide variety of structures from its central commercial district through to its periphery [11,12]. Depending on the area, energy demand patterns such as electricity and car usage are highly variable, owing to a wide variety of social activities [13]. In addition, building structures, materials, usage, and layouts affect heating and cooling demand and PV electricity generation via shading and energy balance [14–16]. These factors need to be considered to adequately evaluate the viability of renewable energy projects. Recent developments in 3D building and energy modeling tools can be used to evaluate these factors [17–20]. "Rhinoceros 3D" is a program used to model 3D building structures combined with GIS data [12]. Using "Grasshopper", a plug-in for the Rhinoceros 3D, solar radiation on building surfaces can be evaluated [16,21]. In addition, energy consumption (e.g., space cooling and heating, appliances, lighting) in buildings can be estimated using other plug-ins (e.g., ladybug, honeybee) that utilize a widely used energy model "EnergyPlus" [15]. These models provide a comprehensive set of analytical tools to evaluate urban energy balances and quantify its environmental conditions [20].

*1.3. Technoeconomic analysis and its application*

To evaluate variable renewable energy (VRE) technologies, such as PV and wind power, it is necessary to conduct technoeconomic analyses [22–24]. A techno-economic model analyzes variable energy generation and demand, and evaluates the financial viability of renewable energy projects with respect to an existing energy system [1,7]. Various financial metrics can be used to evaluate such systems [25,26]. Net Present Value (NPV), which is often used for technoeconomic analyses, compares financial performance against a baseline case, such as the continued use of grid-sourced electricity [3,27]. By evaluating this over a fixed project period and defined discount rate, the financial merit of a project can be determined. Payback periods represent the duration of time required for recovery of upfront investment costs [26]. Internal return rate (IRR) represents the expected annual rate of growth from a project and is defined as the discount rate necessary to have a zero NPV by the end of the project period [25].

Beyond its economic merits, the techno-economic approach allows various financial and sustainability indicators to also be evaluated [2]. For example, reductions in $CO_2$ emissions from a renewable project can be obtained [7]. "Cost saving" indicates how much cost can be saved by installing renewable energy systems compared to the cost of using existing systems. "Self-consumption" represents how much on-site generation can be consumed within the system. "Self-sufficiency" indicates how much demand can be supplied by on-site generation considering a system's demand-supply balance. "Energy sufficiency" indicates the relative difference between annual renewable energy generated on-site and the annual energy demand of the system. With these indices, it is possible to evaluate how a renewable energy project can contribute to the decarbonization of the power system.

The techno-economic analysis can be applied at various spatial scales such as households, urban districts, cities, and countries [28,29]. For the effective use of renewable energy, which are highly variable in time and space, spatial integration of demand and generation becomes critical from household-level to cross-regional scale using grids [30]. It has been indicated in other studies that demand aggregation through microgrid, peer-to-peer exchange, shared storage may enhance economic performances [29,31–33]. Orehounig et al. [32] showed that integrating neighborhood decentralized energy systems (PV, biomass, and small hydro power) with district heating and energy storage can lower demand peaks and consumption. Long et al. [34] indicated that their P2P system in a microgrid can save the community 30 % of the energy cost from conventional grid electricity. However, none of these studies have investigated the cross-scale effects on costs between households and districts, and decarbonization with EV and V2H systems.

Therefore, in this paper, we investigated how PV, battery, and EV systems can contribute to district-scale decarbonization using smart-meter data from a residential area in Shinchi, Fukushima and estimated hourly energy demand data for a central district in Kyoto using Rhinoceros 3D (Table 1). Then, techno-economic analyses were conducted for PV, battery, and EV renewable energy technologies from 2020 to 2040 with declining costs. The following research questions are addressed:

1. How can the application of PV, battery, EV (i.e., V2H system) for 50 individual houses contribute to economic cost saving and decarbonization?
2. How does the aggregation of individual household demand (e.g., microgrids or peer-to-peer exchanges) as a Vehicle to Community (V2C) system, contribute to cost savings and $CO_2$ emission reduction compared to individual applications?

3. How do the effects of PV and EV on cost savings and decarbonization compare between residential and commercial districts, and how do they change over time?

Table 1. Type of districts in Shinchi, Fukushima and Kyoto.

| Location | Shinchi, Fukushima | Kyoto |
|---|---|---|
| Building | 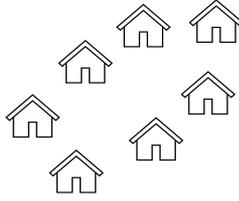 | 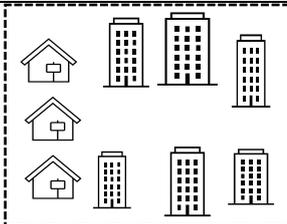 |
| Type | Residential | Commercial |
| Latitude and longitude | 37.9°N, 140.9°E | 35.0 °N, 135.8°E |
| Data type | Smart meter | Rhinoceros 3D |
| District annual demand (kWh) | 295,730 | 2,919,000 |
| District area (m$^2$) | 26,200 | 47,900 |
| Number of building | 50 | 114 |
| Number of EV | 50 | 100 |
| Analysis 1 | Individual houses | N/A |
| Analysis 2 | Aggregated | Aggregated |

This paper is structured as follows: methods, data, and assumptions for the techno-economic analyses are introduced in Section 2. Results of the residential and commercial district analyses are presented in Section 3. The implications of these results for urban decarbonization are discussed in Section 4, and conclusions presented in Section 5.

## 2. Methods, data, and assumptions

*2.1. Techno-economic analyses*

To evaluate the financial viability of renewable energy projects, it is necessary to consider various factors such as PV and battery system costs, profiles of electricity demand and PV generation, levels of insolation, degradation of PV and battery systems, electricity tariffs, project periods, and discount rates [2,3,24,35]. Techno-economic analyses can provide effective insights on the interdependencies and their effect on results. With various PV and battery capacities, techno-economic analyses utilize discounted cashflows by quantifying the financial differences that arise from changes in energy utilization with respect to an existing baseline (i.e., continued use of grid-sourced electricity and a gasoline vehicle). These differences from the baseline are then used to calculate the expected overall cost savings and reductions in CO$_2$ emissions from the renewable energy project. In the "PV + EV" case, the existing systems also include the use of Internal Combustion Engine (ICE) vehicles. As in our previous study [7], for the cost of an EV we only consider the cost difference between EV and ICE purchase prices along with and any additional V2H system costs. The general methodology for this paper follows our earlier studies [2,7,8]. We used 110 yen·$^{-1}$ for the currency exchange rate to USD. A project period of 25 years and discount rate of 3% were used, which are typical values for PV projects in Japan. Following analyses were conducted using Matlab, with results further verified against the System Advisor Model [36].

*2.2. Net present value*

We used NPV as the main financial metric to evaluate different PV capacities with electric or ICE vehicles or battery capacities. The "PV + EV" and "PV (+ battery)" configuration that provide the highest NPV represents the most cost-effective solution in a particular simulation year between 2020 and 2040 [7].

NPV for "PV + EV" can be expressed as:

NPV$_{totoal}$ = NPV$_{electricity}$ + NPV$_{gasoline}$

where

$$NPV_{electricity}(p, b, t) = \sum_{n=1}^{N} \frac{CashFlow(p, b, n, t)}{(1 + R_d)^n} - SysmteCost(p, b, t)$$

$p$ = PV capacity (kWh)
$b$ = Battery capacity (kWh)
$t$ = Project first year (year)
$N$ = Project period (year)
$R_d$ = discount rate

and

Cash Flow ($p, b, n, t$) = Electricity Cost$_{Base}$ ($n, t$) – Electricity Cost$_{System}$ ($p, b, n, t$)

where the subscripts, "Base" and "System" indicate electricity costs (grid electricity) "without" and "with" renewable energy systems (PV, battery, EV) including operation costs and EV battery replacement costs if any, respectively.

$$\text{Electricity Cost}_{Base}(0, 0, n, t) = E_{import}(0, 0, n, t) \cdot T_{import} - \text{Export}(0, 0, n, t) \cdot T_{export}$$

$$\text{Electricity Cost}_{system}(p, b, n, t) = E_{import}(p, b, n, t) \cdot T_{import} - \text{Export}(p, b, n, t) \cdot T_{export} + p \cdot M_{pv} + b \cdot R_{battery}(t)$$

where

$p$ = PV capacity (kW)
$b$ = Battery capacity (kWh)
$E_{import}$ = Electricity imported from grid (kWh yr$^{-1}$)
$E_{export}$ = Electricity exported to grid (kWh yr$^{-1}$)
$T_{import}$ = Flat-rate electricity usage charges ($ kWh$^{-1}$)
$T_{export}$ = Flat-rate feed-in tariff rebate ($ kWh$^{-1}$)
$M_{PV}$ = PV maintenance cost including inverter replacements ($ kWh$^{-1}$ yr$^{-1}$)
$R_{battery}$ = EV battery replacement cost ($ kWh$^{-1}$). Battery is replaced when the capacity degrades to less than 80% of the initial state.

System costs are initial investments to install PV and battery systems.

$$\text{System Cost}(p, b, t) = p \cdot C_{pv}(t) + b \cdot C_{battery}(t)$$

where

$C_{pv}(t)$ = Cost of PV system ($ kW$^{-1}$) in year $t$
$C_{battery}(t)$ = Cost of battery system ($ kWh$^{-1}$) in year $t$

For the "PV + EV" system, $C_{battery}(t)$ ($ kWh$^{-1}$) in year $t$ was calculated as:

$$C_{battery}(t) = EV_{add}(t) / v$$

where

$EV_{add}(t)$ = EV additional cost ($t$) ($/vehicle)
$v$ = EV battery capacity (kWh/vehicle)

$EV_{add}(t)$ is further broken down to:

$$\mathrm{EV_{add}}(t) = \mathrm{EV_{pur}}(t) + C_{\mathrm{V2H}}(t)$$

where $\mathrm{EV_{pur}}(t)$ = Additional costs of EV purchase relative to ICE ($/vehicle) in year $t$, and $C_{\mathrm{V2H}}(t)$ = V2H system cost ($/vehicle) in year $t$.

$\mathrm{NPV_{gasoline}}(n, t)$ represents saving for not using gasoline and was obtained as:

$$NPV_{gasoline}(n,t) = \sum_{n=1}^{N} \frac{Cash\ Flow_{gasoline}(n,t)}{(1+R_d)^n}$$

where Cash Flow$_{\mathrm{gasoline}}$ (n, t) = $t \cdot k \cdot g \cdot u$

$t$ = Total number of passenger vehicles

$k$ = Annual average driving distance (km/vehicle yr$^{-1}$)

$g$ = Gasoline efficiency for ICE (l km$^{-1}$)

$u$ = Unit gasoline price ($ l$^{-1}$)

## 2.3. Financial metrics

We also utilized several financial metrics to assess the viability of projects. Each metric has its own merits and demerits. Depending on the situation and context, different indices are utilized [25]. In this study, several financial metrics are compared to provide a clearer picture of the financial performance of a potential configuration. Simple payback period (SPB) is the time necessary to repay an initial investment cost [26]. This metric is highly intuitive and it is an useful proxy for the decision making of citizen's [25]. However, SPB does not consider the opportunity cost of money and ongoing benefits of PV generation after the payback period. NPV overcomes this weakness of SPB, and many studies use NPV as the basis for their techno-economic analyses. However, NPV requires discount rates to be specified beforehand, but discount rates in practice can vary across different households and context. Real internal return rate (IRR) provides a financial metric without assuming the discount rate (since the IRR is the rate that produces NPV = 0). However, IRR cannot provide information on the scale of considered projects. Therefore, in this study we consider these metrics collectively to draw more robust conclusions on financial performance.

The SPB can be expressed as the duration (e.g., years) from the initial investment to the time when the following condition is satisfied [26].

$$\sum_{n=1}^{t} \Delta I_n \leq \sum_{n=1}^{t} \Delta S_n$$

where *ΔI* is non discounted incremental investment costs and *ΔS* is non discounted sum value of the annual cash flows net annual costs. *t* represents the time when the condition is satisfied for the first time.

The IRR is the rate (*d*) which produces NPV to be zero [26], which can be represented as follows.

$$0 = NPV = \sum_{n=0}^{N}\left(\frac{F_n}{(1+d)^n}\right)$$

where $F_n$ is cash flows received at time *n*, and *d* is the rate that makes *NPV* = 0.

*2.4. Energy indices*

The effectiveness of renewable energy can be assessed using various energy sustainability indices. Energy Sufficiency, ES (*p, b*), self-sufficiency, SS (*p, b*), self-consumption, SC (*p, b*), cost saving, CS (*p, b*) are calculated as:

ES (*p, b*) = $E_{PV}$ (*p, b*) / $E_{load}$ (*p, b*) × 100 (%)
SS (*p, b*) = ($E_{PV\text{-}load}$ (*p, b*) + $E_{battery\text{-}load}$ (*p, b*)) / $E_{load}$ (*p, b*) × 100 (%)
SC (*p, b*) = ($E_{PV\text{-}load}$ (*p, b*) + $E_{battery\text{-}load}$ (*p, b*)) / $E_{PV}$ (*p, b*) × 100 (%)
CS (*p, b*) = $\left\{1 - \frac{NPV_{total}(i,p,b,t)}{N}/\sum_{n=1}^{1} Energy\ Cost_{Base}\ (i,n,t)\right\} \times 100$ (%)

where
$E_{PV}$ (*i, p, b*) = Electricity generated by PV (kWh yr$^{-1}$) in the *i*th house
$E_{load}$ (*i, p, b*) = Total electricity load (kWh yr$^{-1}$) in the *i*th house
$E_{PV\text{-}load}$ (*i, p, b*) = Electricity (kWh yr$^{-1}$) supplied from PV directly to load in the *i*th house
$E_{battery\text{-}load}$ (*i, p, b*) = Electricity (kWh yr$^{-1}$) supplied from battery to load in the *i*th house

*2.5. Assessment of $CO_2$ emissions from the systems*

For our analyses, we considered $CO_2$ emissions from electricity generation and gasoline combustion [7].

$EMI_{Base}$ (*i*) = *e(i)* · $E_{imported}$ + *t(i)* · *k(i)* · *g* · *h*
$EMI_{System}$ (*i*) = *e(j)* · $E_{imported}$

$$\text{CO}_2 \text{ emission reduction (\%)} = (1 - \text{EMI}_{\text{system}}(i) / \text{EMI}_{\text{Base}}) \cdot 100$$

where $\text{EMI}_{\text{Base}}(i)$ is $CO_2$ emission in the Base scenario for the $i$th house, and $\text{EMI}_{\text{System}}(i)$ is $CO_2$ emission in the technology scenarios for the $i$th house. $e(j)$ is grid emission factor ($kg_{CO2}$ $kWh^{-1}$) and $h$ = gasoline emission factor ($kg_{CO2}$ $L^{-1}$) (Table 2).

Table 2. Transport related parameters [8]

| Items | Unit | Numbers | Notes and refs |
|---|---|---|---|
| Gasoline efficiency of private cars (2018) | km·L$^{-1}$ | 12.6 | [8] |
| Gasoline price | $·L$^{-1}$ | 1.29 | Average price of 2008–2018 in Japan [8] |
| $CO_2$ emission from gasoline | $kg_{CO2}$·L$^{-1}$ | 2.3 | [37] |
| $CO_2$ emission from electricity (2018) | $kg_{CO2}$·kW$^{-1}$ | 0.352 for Kyoto<br>0.522 for Shinchi | KEPCO [38]<br>Tohoku [39] |
| Fuel economy of Nissan Leaf (40 kWh; 2019) | km·kWh$^{-1}$ | 5.3 | EPA fuel economy [40] |

*2.6. Electricity tariffs*

Electricity tariffs are highly variable after the full liberalization of the retail sector in 2016 in Japan. However, they can be generally subdivided into two categories, tariffs for "low" and "high" voltages [41]. Low voltage prices are generally for households and high voltages are for industry. We applied low voltage price for the assessment of households in Shinchi, Fukushima. The central district in Kyoto is assessed as a commercial area with a mix of high and low voltage prices [8]. We also considered two additional cases with and without FITs listed in Table 3.

Table 3. Electricity tariffs in Shinchi and Kyoto

| | Import cost | Export value |
|---|---|---|
| Households in Shinchi | 0.22 $·kWh$^{-1}$ | 0.09 $·kWh$^{-1}$ |
| Central district in Kyoto | 0.18 $·kWh$^{-1}$ | 0.08·$·kWh$^{-1}$ |

*2.7. Cost of technologies*

Costs of PV, battery, and EV are expected to drop rapidly for coming decades (Fig. 4) [4–6], and new technologies, such as solid-state batteries [42] and perovskite solar cells [43] are being developed that can increase their urban application potentials, due to such factors as higher efficiencies, lighter weight, and softer materials [14]. Residential rooftop PV system costs (< 10kW) in 2019 were 2.2 $/W

in Japan and is expected to drop to 1.0 $/W by 2030 [44]. Small-scale battery system costs were 1,182 $/W in 2019 [45]. PV plus battery systems are estimated to become more economic than "PV only" systems at a battery price of 636 $/W in Japan (i.e., battery price parity), and this is considered to be the target battery system price for 2030 by the Japanese government [45]. EV costs are calculated as "EV additional costs" which equals the difference between ICE and EV costs with a V2H system [2,46]. We consider Nissan Leaf as the model EV as it has been the bestselling EV in Japan over recent years [47]. Nissan Leaf has a 40kWh battery and cost about $35,000 in 2020 (the middle price from its model variants), and a similar type of ICE (NISSAN Sylphy) is about $22,000. Therefore, the price difference is set at $13,000 in 2020. Nichicon is the most popular V2H system in Japan since 2012. In 2021, the standard V2H model is $4,000, and the premium model is $8,000 with a typical installation cost of around $2,700.

In October 2020, the prime minister of Japan announced that Japan will realize a carbon neutral society by 2050, and that all the new vehicles should be electrified by 2035 (this also includes hybrid vehicles). Since then, various subsidies were introduced for EV and V2H systems. The Ministry of Environment provides a subsidy for EVs to a maximum of $7,300 and conditional on having a 100% renewable electricity supply [48]. The V2H system is also supported by additional subsidies for half of the V2H system price plus the installation cost (maximum $3,600) [48]. It indicates that the Nichion V2H system can be installed by $2,000 and $4,000 for standard and premium models, respectively. Therefore, we estimate the EV additional cost including the V2H system costs is $8,000 in 2021 (EV additional cost = EV-ICE price difference ($13,000) – EV subsidy ($7,000) + V2H system cost ($3,000)). As we assume a Nissan Leaf with a 40 kWh battery capacity, it can be expressed as 200 $/kWh. In 2030, EV price is estimated to be 10-20% cheaper than the ICE price [49], and the V2H system cost is estimated to be around $3,000 including the installation cost, which is similar to the maximum option price at the time of car purchase. Thus, the additional EV cost is estimated to be 25 $/kWh in 2030 [8]. The annual rate of decline from 2020 to 2030 (Table 4) was extended to 2040 to investigate the effects of further cost reductions (Fig. 4).

Table 4. Technology costs estimated for 2020 and 2030 and annual rate of decline.

| Technology component | 2020 | 2030 | Annual rate |
|---|---|---|---|
| Small-scale PV system cost ($/kW) | 2.2 | 1.01 | 0.925 |
| Small-scale battery system cost ($/kWh) | 1,182 | 636 | 0.94 |
| EV additional cost ($/kWh) | 200 | 25 | 0.81 |

*2.8. Electric vehicles as battery*

Vehicle utilization rate is low in Japan especially in urban areas [50–52]. For example, private passenger vehicles are only out-of-home for less than 30 minutes a day on average in urban areas such as Kyoto City [8], and average time of out-of-home duration around Japan is 5.2 hours a day [53]. Due to ongoing cost decreases for PV and EV, high economic values and decarbonization potential can be realized by using an EV battery as energy storage for rooftop PV increasingly so towards 2030 [2,8]. For a single home, the usage pattern of EV can be a bottle neck to maximize the usefulness of "PV + EV" systems [2], but across a community, the sharing of many EV batteries (in a Vehicle to Community (V2C) arrangement) can be used to effectively mitigate the intermittency of individual EV availability and increase the overall availability of EV storage, which then improves the ability of the entire community of district to utilize local rooftop PV generation rather than grid-sourced electricity.

We performed simple experiments to evaluate the effectiveness of community scale EV battery utilization. If a vehicle is out of home for randomly 3 hours a day from 7am to 7pm (daytime), the vehicle is available as a V2H battery for PV for 75% of the daytime on average (Fig. 5). If only one EV is connected to the system, the available battery capacity can become zero for three hours a day (Fig. 5 top). With increasing number of EVs in the system, available battery capacity for V2H during the daytime stabilize to around 75% as each EV is on average out of home for three hours a day (Fig. 5). A bottom panel in figure 5 shows that minimum battery capacity in the system in a day increase from 0 % to above 60% with increasing number of EVs in the system. From around 40 EVs, minimum battery capacity stabilizes around 60 % toward 100 EVs (Fig. 5 bottom). Therefore, with the assumed EV utilization, more than 40 EVs connected to the systems are necessary to stability the system.

In this study, we assumed that 100 EVs are connected to the system for the Kyoto district and 50 EVs are connected for the Shinchi area (Table 1). With more than 40 EVs connected to the system, we can assume that the intermittency of EV availability is largely mitigated. For the analyses, 75% the total EV battery capacity is available for V2C from 7am to 7pm (Fig. 5 bottom). Nissan Leaf is used as the representative EV with a 40 kWh battery. Following our earlier studies [2,7,8], 50% of EV battery is used for V2H, providing the remaining 50% of the battery capacity for driving. Thus, half of 2,000 kWh (40 kWh × 50 EVs) and 4,000 kWh (40 kWh × 100 EVs) of storage is made available for Shinchi and Kyoto communities, respectively.

*2.9. Weather file*

The weather information such as solar insolation and temperature were created using a program "SIREN" [54] that uses reanalysis data (integrated weather data from models and observations), "MERRA-2" [55] for Kyoto and Shinchi. PV capacity factors for Kyoto and Shinchi were calibrated to average observed values for each prefecture [56].

*2.10.     Hourly demand data for residential areas in Shinchi*

Shinchi, Fukushima (38.0°N, 140.6°W) is located north of the main island of Japan facing the Pacific Ocean. The population is around 8,000. Mean annual temperature is around 12°C. Minimum winter and summer monthly average temperatures are around 2 °C and 24 °C, respectively. This region has limited snow in winter. Shinchi is a small town in Fukushima prefecture. The number of private passenger vehicle is 6,700 for 8,250 people in 2016, indicating 0.81 vehicles per capita. As average number of people per household is 3.2 in Shinchi. Average number of vehicles per household is 2.6.

We used smart meter data for 50 individual houses in Shinchi, Fukushima for 2018 (Fig. 1 and 2) [57]. These data were nearly continuous for 8760 hours for 2018. Sporadic hourly data were still lacking, and we filled them using averages of the data of the same hour of consecutive 3 days including the day of data deficiency. The basic statistics of these hourly demand data are given in Table 5. In Shinchi, peaks for winter heating demands are larger than the peaks for summer cooling demands as it is in northern Japan.

In our study, we assumed one of the vehicles was converted to EV and each household installed a V2H system. For the analysis of 50 individual houses in Shinchi, Fukushima, we assumed that EVs are out of home for three hours a day. In the 12 hours between 7am to 7pm, three sets of "one-hour blocks" were randomly selected. When the car is out of home, EV charging/discharging is not possible. For each 1-hour trip, 1.1 kWh (5.8-km drive) of electricity is consumed from the EV battery. This corresponds to 6,368 km/year of driving. At 12 am, if the state of the charge (SoC) of the EV battery is less than 50%, the battery is grid charged to a SoC of 50%.

Table 5. Statistics of hourly demand data from Shinchi, Fukushima

| Items | Numbers |
|---|---|
| Period | Jan 1, 2018 to Dec. 31, 2018 |
| Number of houses | 50 |
| Resolution | Hourly |
| Mean of annual demand (kWh) | 5,915 |
| Maximum of annual demand (kWh) | 13,894 |
| Minimum of annual demand (kWh) | 1,511 |
| Hourly maximum demand (kW) | 12.3 |
| Hourly minimum demand (kW) | 0.05 |

*2.11.    Hourly demand modeling of a central district in Kyoto using Rhinoceros 3D*

The hourly demand data was also estimated utilizing the 3D energy balance model, "EnergyPlus", as a plug-in for Rhinoceros 3D (Fig. 2), which is increasingly used for building energy analyses [15,17]. In cities, neighboring buildings affect the energy balance of other buildings depending on their layouts. Shading from neighboring buildings affect a building's available solar radiation and subsequent PV generation, while also influencing its cooling and heating demand. These effects can be simulated using Rhinoceros 3D and its plug-ins, "Grasshopper" [17,18,21]. Using these programs, it is possible to analyze rooftop radiation potential and the energy consumption of buildings with various spatial and temporal resolutions (hourly in this study). In our analysis, the 3D building structures were reconstructed as polygons using the footprint data of buildings and their respective heights (Fig. 3) [8,58].

Hourly electricity demand over 2018 was estimated for the central commercial district in Kyoto City using a weather file. Kyoto City (35.0°N, 135.7°W) has an annual average temperature of around 16 °C. Minimum winter and summer monthly average temperatures are around 5 °C and 28 °C, respectively. 114 buildings were identified for the Kyoto central district in an aera of 47,900 $m^2$ (Table 1, Fig. 3). Total rooftop area is calculated to be 26,952 $m^2$, which corresponds to 56% the total area for the district. For a comparison, we analyzed a block of typical households with 43 homes in Shinchi, Fukushima (Fig. 3). The district has a total area of 26,200 $m^2$, and the toral rooftop area of 43 houses 6,600 $m^2$ with an average of 153 $m^2$, corresponding to 25% of the total area. The Kyoto central district is denser with taller buildings compared to Shinchi, indicating that each building in the Kyoto district is more likely to affect the PV generation and energy balance of neighboring buildings. In the Kyoto district, the rooftop area that has minor shading from neighboring buildings (i.e., >95% of full radiation) is 47% of the total rooftop area. Rooftop area receiving >70% of full radiation is 84% of the total rooftop area (Fig. 3).

To estimate building energy demands, we need to know the types of activities that are taking place in the spaces within these buildings. Buildings in the Kyoto district are primarily used for commercial applications (shops, restaurants, offices, hotels, and residences), and in Shinchi they are all residential. In the model, we assumed that the central district in Kyoto has all commercial buildings with concrete materials, and Shinchi residential houses are built using wooden materials. We select the closest options in the program (EnergyPlus in Rhinoceros 3D), namely "supermarket" and "mid-rise apartment" for Kyoto and Shinchi, respectively. According to these specifications, the energy demands (e.g., lighting, electric equipment, and HVAC) were calculated for each building consistent with the energy balance through building structures under changing solar radiation and temperatures (Fig. 2). For commercial building in the Kyoto district, daytime electricity consumption is relatively stable with a small peak in the morning for space heating. For the residential district in Shinchi, two large peaks occur in the morning and the evening. The late night to morning peak is larger in the observed data, which is likely due to electric thermal storage heaters that are used in the region and timed for the usage of cheap nighttime electricity (e.g., 23:00-7:00) (Fig. 2).

Total annual electricity demand for the areas in Kyoto and Shinchi are determined to be 2,863,605 kWh and 509,807 kWh respectively (Table 1), including heating, cooling, lighting, and interior equipment. Compared with the observed smart meter data for Shinchi, the modelled data for an average household is two times larger possibly owing to house sizes (the second floor is often smaller than the first floor in the region, but the first and second floors have the same shape in the model (Fig. 3)). When the estimated demand is divided by two, the observation and modelled data agree well indicating seasonal and daily variability in heating and cooling are well captured in the model ($r = 0.95$; Fig. 4). For Kyoto, the modelled demand data are used with no correction for the following technoeconomic analyses. Although the model may overestimate the demand of the Kyoto district, the purpose of this study is the relative comparison between Kyoto and Shinchi to show the difference of economic viability between high and low demand districts, therefore this potential overestimate issue is not a critical problem.

For following analyses, we used observed smart meter data for Shinchi and simulated hourly demand estimates for Kyoto.

## 3. Results

### 3.1. Scenarios

We analyzed and compared two technology combinations, "PV (+ battery)" and "PV + EV" systems from 2020 to 2040. For each scenario, ranges of PV and battery capacities were analyzed, and the configurations with the highest NPV were chosen to represent the scenario [2,8]. As is shown later, the battery does not have an economic advantage until after 2030 in Japan. In the case of Shinchi, two scenarios of individual analyses for 50 houses and aggregated analyses for summed demand were analyzed. For Kyoto, only aggregated analyses are conducted as the random variability for each building was not modeled. The results of aggregated analyses for Shinchi and Kyoto are compared to illustrate the difference in technology potentials under different urban environments. In addition, cases with and without FITs were analyzed.

### 3.2. Individual houses and aggregated district analyses for Shinchi

We conducted two sets of technoeconomic analyses for Shinchi houses from 2020 to 2040. First, hourly electricity demands of 50 houses were individually analyzed. Second, the individual data is aggregated and analyzed. The average values of individual analyses are presented in figure 6 and 7 (Table S1 in Appendix) with the results for the aggregated analysis. The results of aggregated analyses are divided by 50 so that the results are comparable with the results of individual analysis. The aggregated analyses assume that excess electricity can be exchanged perfectly between houses within the residential district (e.g., microgrid behind the meter).

PV capacity shows individual and aggregated analyses have nearly the same optimum values for all the cases (Fig. 6). With FIT, optimum PV capacity reaches its maximum (10 kW) for all the cases around 2026. Aggregated analyses reach the 10 kW 1-year earlier reflecting better economics with larger PV capacity than that of the individual cases. It is noted that "PV + EV" cases increase optimum PV capacity about four times more than "PV (+ battery)" cases without FIT case. For the "PV (+

battery)" cases, batteries become economic only after 2036 in the "without FIT" cases, indicating that "PV (+ battery)" cases are mostly "PV only" system during the analysis period. For the "PV + EV" cases, available battery capacities are set to be half of full 40 kWh (20 kWh) for each home (Fig. 6).

Regarding the range of NPVs, "PV + EV" and "with FIT" cases generally show larger NPVs compared to "PV (+ battery)" and "without FIT", respectively, which are consistent with our earlier study [2]. "PV (+ battery)" cases show that NPVs are indistinguishable for aggregated and individual analyses (Fig. 6). On the other hand, "PV + EV" cases increase NPVs significantly for aggregated analyses than individual cases, and more so for "without FIT" case (Fig. 6). For example, in 2030, aggregation increases NPV of individual analyses of "PV + EV without FIT" case by 23 %. The results indicate that electricity exchanges between houses are important for "PV + EV" cases to increase the benefits because of the large battery capacity. It is noted that NPVs for "PV (+ battery)" are larger than that of "PV + EV" around 2020 owing to the higher costs of the EV additional costs.

"PV + EV" increase self-consumption substantially (Fig. 7). Even with maximum PV capacity (10 kW), self-consumption increases to 48% ("PV + EV" with FIT) from 17% ("PV (+ battery)" with FIT). Aggregated analyses further increase self-consumption by 11 points (meaning 17 % to 28 %) indicating aggregated "PV + EV" system is 3.5 times more efficient for consuming PV electricity than the PV only system.

Self-sufficiency indicates high values for "PV+EV" cases of 65 % to 98 % (Fig. 7). Aggregation of the demands increase the self-sufficiency by 12 points in percentage on average. With FIT case, "PV+EV" can provide nearly 100% of electricity from the system. "PV (+ battery)" provides only up to 40% of self-sufficiency. It is noted that self-sufficiency of "PV (+ battery)" without FIT case slightly increases owing to the addition battery in late $2030^{th}$ (Fig. 6).

Energy sufficiency increases up to 200% when FIT is available owing to larger PV capacities (Fig. 7). The difference of energy sufficiency between aggregated and individual analyses with FIT originate only mathematical reasons as the total PV capacities and demands for aggregated and individual analyses are the same for "with FIT" cases. This is caused by the fact that energy sufficiency calculation is not linear (small demands as denominators with the same PV generation in households create larger summed energy sufficiency than that of the aggregated analysis).

$CO_2$ emission reduction for "PV + EV" cases are much larger than "PV (+ battery)" owing to more $CO_2$ free PV electricity consumption plus no gasoline combustion. As most of the electricity demand can be supplied from the "PV + EV" system (high self-sufficiency), household $CO_2$ emissions from electricity and vehicle use can be mostly eliminated up to 94 % for aggregated analyses. Aggregation of demand can contribute to $CO_2$ emission reduction of about 6-point increase in percentage on average as self-consumption increases by aggregation.

Changes in cost saving show a cross over point around 2020, when the cost savings for "PV + EV" becomes larger than that of "PV (+ battery)" owing to the rapid cost decline of PV and EV. "PV + EV" without FIT shows larger difference in individual and aggregated analyses as the EV battery plays a more important role when excess electricity cannot be sold to grid. "PV + EV" have larger cost saving by 22 points in percentage than that of "PV (+ battery)" in 2030. Aggregated "PV+EV" analysis without FIT in 2030 shows 5 points in percentage higher cost saving than that of individual analyses.

Simple payback period generally decreases from 2020 to 2040 except a period of PV capacity expansion in PV (+ battery) with FIT cases around 2025 and battery addition around late 2030 (Fig. 6 and 7). Although "PV + EV" had longer payback periods of 14-15 year than that (11-13 years) of "PV (+ battery)" in early 2020s, it reduces to 5-6 years, which is smaller than that (7-10 years) of "PV (+ battery)" cases.

IRR also shows a similar trend with "PV + EV" being not economic (2-4%) around 2020 compared with that (7-9 %) of "PV (+ battery)", but by around 2025, IRRs for "PV + EV" become larger than that of "PV (+ battery)". Then, IRR for a "PV + EV" case reaches 18% on average in 2030, which is larger than that (12%) of "PV (+ battery)".

### 3.3. Central district in Kyoto

In comparison to Shinchi, the central district in Kyoto has 114 taller buildings of various sizes and are closely packed with limited number of parking spaces (Fig. 3). The analyses were conducted using estimated hourly demand profile by Rhinoceros 3D and 100 EVs. All the buildings were assumed to be used for commercial building purposes in the model (Fig. 2). Although we analyzed individual buildings for heating, cooling, electrical appliances, and lighting considering shading influences of neighboring building, we conducted techno-economic analyses on aggregated district demand because random variabilities of energy consumption in each building were not modelled.

PV capacity reaches the maximum value around 2027 for the cases with FIT (Fig. 8; Table S2 in Appendix). Without FIT, the PV capacity grows slowly from 2020 to 2040 reflecting gradual cost PV declines. The merit of "PV + EV" in comparison to "PV (+ battery)" cases is smaller than that of the Shinchi cases as Kyoto district has limited number of EVs with regard to its size of its energy demand.

NPV analysis also shows similar characteristics of increasing trends, and higher values for "PV + EV" and "with FIT" cases than "PV (+ battery)" and "without FIT". "PV + EV" without FIT and "PV (+ battery)" with FIT cases have similar trends over the period.

Self-consumption, self-sufficiency, energy sufficiency also shows similar trends with the Shinchi case. The slight difference in energy sufficiency for "PV + EV" and "PV (+ battery)" with FIT after 2028 is because of the electricity losses due to the use of EV batteries.

High $CO_2$ emission reduction of up to 74% can be obtained by "PV + EV" than that (up to 51%) of "PV (+ battery)".

Cost savings also show larger increases by combining with EV than that of the "PV (+ battery)" cases, reaching 24 % on average for "PV + EV" and 16 % for "PV (+ battery)" in 2030, respectively.

Payback period declines from 14 years in 2020 to 7 to 8 years in 2030 with the "PV + EV" cases showing shorter payback periods than that of "PV (+ battery)" by 1 year in 2030. Smaller PV capacities in "without FIT" cases show smaller payback period that that of "with FIT" cases.

IRR starts from 5-6 % and reaches 10 to 16 % in 2030, showing 2 % higher IRR in "PV + EV".

*3.4. Comparison between Shinchi (residential) and Kyoto (commercial) districts*

Represented by Shinchi and Kyoto districts, we compare different urban areas (i.e., residential and commercial) in terms of rooftop area, number of EVs, size of demand, which affect the economics of "PV + EV" and "PV (+ battery)" systems. These results provide important information for city officers and business planning for the decarbonization of urban environments over the coming decades. In figure 9, averages on "with FIT" and "without FIT" for aggregated data are plotted to illustrate differences between the urban areas and the technology combinations.

For self-consumption, "PV + EV" cases are higher than "PV (+ battery)" cases regardless of the area after 2025 (Fig. 9; Table S3 in Appendix). The Kyoto district shows higher self-consumption owing to relatively smaller PV capacities (rooftop area) compared with demand.

For self-sufficiency, "PV + EV" for Shinchi has higher self-sufficiency than that of Kyoto because more PV and EV capacities are available relative to demand (Fig. 9). Self-sufficiency reaches more than 90% for "PV + EV" in Shinchi, indicating high resilience on the time of disaster. The difference between "PV + EV" and "PV (+ battery)" in Kyoto is smaller than that of Shinchi, indicating that adding EV on PV system has smaller contribution in the central urban area with limited parking space.

Energy sufficiency indicates PV can generate electricity more than demand in Shinchi in a year, but it is about 90% in Kyoto. High reductions in $CO_2$ emissions are possible using "PV + EV" cases, reaching 92% in Shinchi and 70% in Kyoto.

Cost saving increasing over time for all the cases and is larger using "PV + EV" compared to "PV (+ battery)". "PV + EV" for Shinchi has a larger cost saving than Kyoto due to the larger PV and EV battery capacity potentials in the residential area, reaching 35% and 24 % in 2030, respectively.

Payback periods of "PV + EV" system reaches 6 years by 2030 in Shinchi and 8 years in Kyoto. Payback periods (9 years) of "PV (+ battery)" is higher in 2030 that that of "PV + EV".

IRR for "PV + EV" in Shinchi is the highest, reaching around 19% in 2030, indicating a growing business opportunity. However, IRR of "PV (+ battery)" in 2020 remains higher than that of "PV + EV" with the value of 6-8%.

Importantly, the financial benefits of using a "PV + EV" system more rapidly improves in Shinchi than in Kyoto as Shinchi has a relatively larger number of EVs. Apparently, economical superiority of the "PV + EV" system changed its location from Kyoto to Shinchi around 2020. In addition, "PV + EV" start showing better financial returns than "PV (+ batter)" in the early 2020s. Therefore, different urban areas and technology combinations changes its economic superiority of its location/technology with time as the costs of PV and EV decreases.

## 4. Discussions and outlooks

Decarbonization has become a central global objective fueled by climate change and declining costs of renewable energy technologies [59]. Urban decarbonization is a critical part of this process as more people move from rural to urban areas [14]. In our paper, we demonstrated that combinations of PV, battery, and EV technologies in different city districts have different decarbonization potentials and cost-effectiveness, that continue to change over time. It is important to consider these influencing factors as city officers and businesses plan for urban decarbonization.

Various uncertainties should be recognized for the analyses [2,8]. An important uncertainty is future costs of PV and battery systems and EVs. Considering a long-term declining trend of these technologies, its decreases over the coming decades are likely but the actual rate of the change could be faster or slower. Therefore, the timing of the realization of estimated costs requires some caution. In addition, we assumed constant tariff, climate, demand, discount rates, etc. These factors affect various scenarios of technologies equally. Therefore, relative changes between technologies are more robust results than those of the absolute values. For example, our analyses indicate that "PV + EV" will achieve greater economic returns over "PV (+ battery)" systems in a few years, and the economic gains using "PV + EV" systems in the residential area of Shinchi will be more economic than the commercial district of Kyoto. As a conclusion, "PV + EV" in residential areas could play a substantial role in urban decarbonization over the coming decades.

"PV + EV" systems have seen rapid development and commercialization in Japan after the Fukushima Daiichi nuclear disaster. However, the markets for V2H systems in other parts of the world have yet to be established. This may be related to car utilization rates. In Japan, the utilization rates of cars are quite low, in particular in suburban regions. Also, Japan is disaster prone region, which promotes decentralized energy systems with large battery capacities. With increasing penetration of rooftop PV and EV in societies, it is likely that the establishment of "PV + EV" systems will become easier around the world.

The successful establishment of "PV+EV" system (SolarEV City) requires smooth electricity exchange between buildings through peer-to-peer (P2P) energy transaction. Although P2P has been extensively studied and many demonstration projects are currently on-going, it has not been fully operational in any parts of the world with suggested reasons such as regulations [30]. Therefore, to realize city-scale application of "PV + EV" requires these regulatory frameworks to first be established by working with regulators, policy makers, industries, and communities.

In our analysis, we considered only $CO_2$ emission from electricity generation and gasoline combustion. However, the production process of EV batteries is known to emit a large amount of $CO_2$ if it uses electricity generated from fossil fuels. A city-scale PV + EV system can supply $CO_2$-free electricity for EV battery production if it is extended to the EV production process. In future studies, it would be useful to consider the impacts of $CO_2$ emission for the lifetime of PV and EV, which could influence purchasing decisions of owners that favor low lifecycle emissions EV and PV systems.

In our study, we only analyzed residential and commercial districts. In future studies, city-wide analysis should be conducted allowing excess-electricity transfer from the suburban residential areas (high PV generation and low demand) to the central commercial district (high demand and low PV generation) while taking grid constraints into consideration.

## 5. Conclusions

We analyzed smart meter data for 50 households in Shinchi, Fukushima for "PV + EV" and "PV (+ battery)" from 2020 to 2040. Individual analyses were compared with aggregated analyses that assumed electricity was freely exchanged within community (e.g., behind-the-meter in a microgrid). Results show that aggregated cases increased cost savings by up to 5 points in percentage in 2030 and $CO_2$ emission reduction also increased by 5 points in percentage, emphasizing the need to develop decentralized energy system such as peer-to-peer trading. We also conducted analyses on a central commercial district in Kyoto City. For Kyoto, we applied 3D building energy modeling (Rhinoceros 3D and its plug-ins) to determine the building electricity demand (cooling and heating, lighting, and appliances) that considered the influence of neighboring buildings on its energy balance and solar radiation. The modeling results are utilized to perform technoeconomic analyses for technology combinations of PV, battery, and EV. The comparisons between Shinchi (residential) and Kyoto (central district) show that "PV + EV" systems were a more effective option, in particular for residential decarbonization with larger cost saving than that of "PV (+ battery)". These results have important implications for how urban decarbonization should be conducted in the coming decades.

**Acknowledgments:** We thank Mr. Konosuke Yabuta at Nichicon for providing information on V2H systems.

**Contribution** Takuro Kobashi: Conceptualization, Methodology, Software, Validation, Formal analysis, Investigation, Resources, Writing - Original, Writing - Review & Editing, Visualization, Supervision, Project administration. Younghun Choi: Formal analysis. Yujiro Hirano: Data Curation. Yoshiki Yamagata: Funding acquisition. Kelvin Say: Writing - Review & Editing, Validation.

# Appendix

Table S1. Results for Shinchi aggregated and individual analyses (Fig. 6 and 7). Indiv. and Agg. Indicate individual and aggregated analyses, respectively.

| Items | Condition 1 | Condition 2 | Condition 3 | 2020 | 2025 | 2030 | 2040 |
|---|---|---|---|---|---|---|---|
| PV cost ($/W) | All | All | All | 2.2 | 1.5 | 1.0 | 0.5 |
| Battery cost ($/kWh) | All | All | All | 1182 | 867 | 637 | 343 |
| EV additional ($/kWh) | All | All | All | 200 | 70 | 24 | 3 |
| PV capacity (kW) | PV (+battery) | w/ FIT | Indiv. | 1 | 1 | 2 | 3 |
| | PV (+battery) | w FIT | indiv. | 2 | 5 | 10 | 10 |
| | PV (+battery) | w/ FIT | Agg. | 1 | 1 | 2 | 2 |
| | PV (+battery) | w FIT | Agg. | 2 | 5 | 10 | 10 |
| | PV + EV | w/ FIT | Indiv. | 5 | 5 | 6 | 7 |
| | PV + EV | w FIT | Indiv. | 6 | 9 | 10 | 10 |
| | PV + EV | w/ FIT | Agg. | 5 | 5 | 6 | 7 |
| | PV + EV | w FIT | Agg. | 6 | 10 | 10 | 10 |
| Battery capacity (kWh) | PV (+battery) | w/ FIT | Indiv. | 0 | 0 | 0 | 6 |
| | PV (+battery) | w FIT | indiv. | 0 | 0 | 0 | 0 |
| | PV (+battery) | w/ FIT | Agg. | 0 | 0 | 0 | 5 |
| | PV (+battery) | w FIT | Agg. | 0 | 0 | 0 | 0 |
| | PV + EV | w/ FIT | Indiv. | 20 | 20 | 20 | 20 |
| | PV + EV | w FIT | Indiv. | 20 | 20 | 20 | 20 |
| | PV + EV | w/ FIT | Agg. | 20 | 20 | 20 | 20 |
| | PV + EV | w FIT | Agg. | 20 | 20 | 20 | 20 |
| NPV ($) | PV (+battery) | w/ FIT | Indiv. | 1,004 | 1,897 | 2,604 | 3,836 |
| | PV (+battery) | w FIT | indiv. | 1,609 | 3,424 | 7,767 | 13,229 |
| | PV (+battery) | w/ FIT | Agg. | 1,480 | 2,294 | 2,963 | 4,049 |
| | PV (+battery) | w FIT | Agg. | 1,821 | 3,523 | 7,906 | 13,368 |
|  | PV + EV | w/ FIT | Indiv. | -1,395 | 7,439 | 12,039 | 16,461 |
|  | PV + EV | w FIT | Indiv. | 631 | 10,967 | 17,434 | 23,752 |
|  | PV + EV | w/ FIT | Agg. | 1,549 | 10,317 | 14,826 | 19,197 |
|  | PV + EV | w FIT | Agg. | 2,187 | 12,588 | 19,207 | 25,523 |
| Self-consumption (%) | PV (+battery) | w/ FIT | Indiv. | 77 | 74 | 70 | 70 |
|  | PV (+battery) | w FIT | indiv. | 68 | 33 | 17 | 17 |
|  | PV (+battery) | w/ FIT | Agg. | 93 | 87 | 75 | 74 |
|  | PV (+battery) | w FIT | Agg. | 75 | 35 | 18 | 18 |
|  | PV + EV | w/ FIT | Indiv. | 78 | 74 | 71 | 66 |
|  | PV + EV | w FIT | Indiv. | 71 | 56 | 48 | 48 |
|  | PV + EV | w/ FIT | Agg. | 96 | 93 | 87 | 79 |
|  | PV + EV | w FIT | Agg. | 87 | 62 | 59 | 59 |
| Self-sufficiency (%) | PV (+battery) | w/ FIT | Indiv. | 21 | 22 | 23 | 40 |
|  | PV (+battery) | w FIT | indiv. | 24 | 33 | 37 | 37 |
|  | PV (+battery) | w/ FIT | Agg. | 19 | 21 | 24 | 35 |
|  | PV (+battery) | w FIT | Agg. | 24 | 32 | 35 | 35 |
|  | PV + EV | w/ FIT | Indiv. | 65 | 71 | 75 | 80 |
|  | PV + EV | w FIT | Indiv. | 76 | 85 | 86 | 86 |
|  | PV + EV | w/ FIT | Agg. | 77 | 82 | 87 | 92 |
|  | PV + EV | w FIT | Agg. | 87 | 97 | 98 | 98 |
| Energy sufficiency (%) | PV (+battery) | w/ FIT | Indiv. | 29 | 32 | 35 | 57 |
|  | PV (+battery) | w FIT | indiv. | 36 | 99 | 271 | 271 |
|  | PV (+battery) | w/ FIT | Agg. | 20 | 24 | 32 | 48 |
|  | PV (+battery) | w FIT | Agg. | 32 | 92 | 200 | 200 |
|  | PV + EV | w/ FIT | Indiv. | 83 | 96 | 106 | 122 |
|  | PV + EV | w FIT | Indiv. | 107 | 158 | 201 | 201 |
|  | PV + EV | w/ FIT | Agg. | 80 | 88 | 100 | 117 |
|  | PV + EV | w FIT | Agg. | 100 | 158 | 166 | 166 |
| $CO_2$ emission reduction (%) | PV (+battery) | w/ FIT | Indiv. | 14 | 15 | 16 | 27 |
|  | PV (+battery) | w FIT | indiv. | 16 | 22 | 25 | 25 |
|  | PV (+battery) | w/ FIT | Agg. | 13 | 15 | 17 | 26 |
|  | PV (+battery) | w FIT | Agg. | 17 | 23 | 26 | 26 |
|  | PV + EV | w/ FIT | Indiv. | 71 | 77 | 80 | 84 |
|  | PV + EV | w FIT | Indiv. | 80 | 88 | 89 | 89 |
|  | PV + EV | w/ FIT | Agg. | 78 | 81 | 85 | 89 |
|  | PV + EV | w FIT | Agg. | 85 | 94 | 94 | 94 |
| Cost saving | PV (+battery) | w/ FIT | Indiv. | 2 | 4 | 5 | 7 |
| (%) | PV (+battery) | w FIT | indiv. | 3 | 7 | 16 | 29 |
| | PV (+battery) | w/ FIT | Agg. | 3 | 5 | 6 | 8 |
| | PV (+battery) | w FIT | Agg. | 4 | 7 | 16 | 27 |
| | PV + EV | w/ FIT | Indiv. | -4 | 15 | 25 | 34 |
| | PV + EV | w FIT | Indiv. | 0 | 23 | 37 | 52 |
| | PV + EV | w/ FIT | Agg. | 3 | 21 | 30 | 39 |
| | PV + EV | w FIT | Agg. | 4 | 26 | 39 | 52 |
| Simple payback period | PV (+battery) | w/ FIT | Indiv. | 13 | 10 | 7 | 8 |
| (year) | PV (+battery) | w FIT | indiv. | 12 | 12 | 10 | 5 |
| | PV (+battery) | w/ FIT | Agg. | 11 | 8 | 7 | 7 |
| | PV (+battery) | w FIT | Agg. | 12 | 12 | 10 | 5 |
| | PV + EV | w/ FIT | Indiv. | 15 | 9 | 6 | 3 |
| | PV + EV | w FIT | Indiv. | 15 | 9 | 6 | 3 |
| | PV + EV | w/ FIT | Agg. | 14 | 8 | 5 | 3 |
| | PV + EV | w FIT | Agg. | 14 | 9 | 6 | 3 |
| IRR | PV (+battery) | w/ FIT | Indiv. | 6 | 10 | 15 | 12 |
| (%) | PV (+battery) | w FIT | indiv. | 7 | 7 | 9 | 23 |
| | PV (+battery) | w/ FIT | Agg. | 9 | 13 | 16 | 14 |
| | PV (+battery) | w FIT | Agg. | 7 | 7 | 9 | 23 |
| | PV + EV | w/ FIT | Indiv. | 2 | 9 | 17 | 42 |
| | PV + EV | w FIT | Indiv. | 3 | 9 | 16 | 40 |
| | PV + EV | w/ FIT | Agg. | 4 | 12 | 20 | 46 |
| | PV + EV | w FIT | Agg. | 4 | 10 | 17 | 43 |

Table S2. Results for aggregated Kyoto District (Fig. 8).

| Items | Condition 1 | Condition 2 | 2020 | 2025 | 2030 | 2040 |
|---|---|---|---|---|---|---|
| PV capacity (kW) | PV (+battery) | w/ FIT | 846 | 1,106 | 1,301 | 1,626 |
| | PV (+battery) | w FIT | 1,041 | 1,886 | 3,252 | 3,252 |
| | PV + EV | w/ FIT | 1,365 | 1,593 | 1,820 | 2,178 |
| | PV + EV | w FIT | 1,560 | 2,405 | 3,250 | 3,250 |
| Battery capacity (kWh) | PV (+battery) | w/ FIT | 0 | 0 | 0 | 0 |
| | PV (+battery) | w FIT | 0 | 0 | 0 | 0 |
| | PV + EV | w/ FIT | 2,000 | 2,000 | 2,000 | 2,000 |
| | PV + EV | w FIT | 2,000 | 2,000 | 2,000 | 2,000 |
| NPV ($) | PV (+battery) | w/ FIT | 587,135 | 1,269,772 | 1,840,988 | 2,628,528 |
| | PV (+battery) | w FIT | 680,839 | 1,662,633 | 2,991,938 | 4,768,265 |
| | PV + EV | w/ FIT | 513,288 | 2,083,097 | 3,084,465 | 4,249,225 |
| | PV + EV | w FIT | 587,826 | 2,450,481 | 4,100,807 | 5,961,479 |
| Self-consumption (%) | PV (+battery) | w/ FIT | 96 | 89 | 83 | 74 |
| | PV (+battery) | w FIT | 91 | 67 | 45 | 45 |
| | PV + EV | w/ FIT | 98 | 94 | 89 | 80 |
| | PV + EV | w FIT | 95 | 75 | 60 | 60 |
| Self-sufficiency (%) | PV (+battery) | w/ FIT | 32 | 38 | 42 | 47 |
| | PV (+battery) | w FIT | 37 | 49 | 56 | 56 |
| | PV + EV | w/ FIT | 50 | 56 | 60 | 65 |
| | PV + EV | w FIT | 55 | 67 | 73 | 73 |
| Energy sufficiency (%) | PV (+battery) | w/ FIT | 33 | 43 | 51 | 63 |
| | PV (+battery) | w FIT | 40 | 73 | 126 | 126 |
| | PV + EV | w/ FIT | 51 | 59 | 68 | 81 |
| | PV + EV | w FIT | 58 | 90 | 121 | 121 |
| $CO_2$ emission reduction (%) | PV (+battery) | w/ FIT | 28 | 34 | 38 | 42 |
| | PV (+battery) | w FIT | 33 | 44 | 51 | 51 |
| | PV + EV | w/ FIT | 53 | 58 | 62 | 67 |
| | PV + EV | w FIT | 57 | 69 | 74 | 74 |
| Cost saving (%) | PV (+battery) | w/ FIT | 4 | 9 | 12 | 18 |
| | PV (+battery) | w FIT | 5 | 11 | 20 | 32 |
| | PV + EV | w/ FIT | 3 | 14 | 21 | 29 |
| | PV + EV | w FIT | 4 | 17 | 28 | 40 |
| Simple payback period (year) | PV (+battery) | w/ FIT | 13 | 10 | 8 | 4 |
| | PV (+battery) | w FIT | 14 | 11 | 9 | 5 |
|  | PV + EV | w/ FIT | 14 | 9 | 7 | 4 |
|  | PV + EV | w FIT | 14 | 10 | 8 | 4 |
| IRR | PV (+battery) | w/ FIT | 6 | 9 | 14 | 26 |
| (%) | PV (+battery) | w FIT | 6 | 8 | 10 | 25 |
|  | PV + EV | w/ FIT | 5 | 10 | 16 | 33 |
|  | PV + EV | w FIT | 5 | 9 | 13 | 31 |

Table S3. Comparison between Kyoto and Shinchi districts aggregated (Fig 9).

| Items | Condition 1 | Condition 2 | 2020 | 2025 | 2030 | 2040 |
|---|---|---|---|---|---|---|
| Self-consumption (%) | Kyoto | PV (+battery) | 94 | 78 | 64 | 59 |
| | Shinchi | PV (+battery) | 84 | 61 | 46 | 46 |
| | Kyoto | PV + EV | 96 | 85 | 75 | 70 |
| | Shinchi | PV + EV | 92 | 77 | 73 | 69 |
| Self-sufficiency (%) | Kyoto | PV (+battery) | 34 | 44 | 49 | 51 |
| | Shinchi | PV (+battery) | 21 | 26 | 30 | 35 |
| | Kyoto | PV + EV | 53 | 62 | 67 | 69 |
| | Shinchi | PV + EV | 82 | 89 | 92 | 95 |
| Energy sufficiency (%) | Kyoto | PV (+battery) | 37 | 58 | 88 | 95 |
| | Shinchi | PV (+battery) | 26 | 58 | 116 | 124 |
| | Kyoto | PV + EV | 55 | 75 | 95 | 101 |
| | Shinchi | PV + EV | 90 | 123 | 133 | 141 |
| $CO_2$ emission reduction (%) | Kyoto | PV (+battery) | 31 | 39 | 44 | 46 |
| | Shinchi | PV (+battery) | 15 | 19 | 22 | 26 |
| | Kyoto | PV + EV | 55 | 63 | 68 | 70 |
| | Shinchi | PV + EV | 81 | 88 | 90 | 92 |
| Cost saving (%) | Kyoto | PV (+battery) | 4 | 10 | 16 | 25 |
| | Shinchi | PV (+battery) | 3 | 6 | 11 | 18 |
| | Kyoto | PV + EV | 4 | 15 | 24 | 35 |
| | Shinchi | PV + EV | 4 | 23 | 35 | 46 |
| Simple payback period (year) | Kyoto | PV (+battery) | 14 | 11 | 9 | 5 |
| | Shinchi | PV (+battery) | 12 | 10 | 9 | 6 |
| | Kyoto | PV + EV | 14 | 10 | 8 | 4 |
| | Shinchi | PV + EV | 14 | 9 | 6 | 3 |
| IRR (%) | Kyoto | PV (+battery) | 6 | 9 | 12 | 26 |
| | Shinchi | PV (+battery) | 8 | 10 | 13 | 19 |
| | Kyoto | PV + EV | 5 | 9 | 14 | 32 |
| | Shinchi | PV + EV | 4 | 11 | 19 | 44 |

Figure Caption

Figure 1. Average hourly demand data for 50 houses in Shinchi, Fukushima for 2018.

Figure 2. Daily electricity demand in Shinchi, Fukushima and Kyoto for 2018. Top panel is observed and modelled average daily demand in Shinchi from household. Bottom panel is modelled daily demand for the Kyoto district. Small windows show average hourly demands in a day for the year.

Figure 3. PV potentials (kWh/m$^2$) on building surface areas. (a) central district of Kyoto City, (b) residential houses in Shinchi, Fukushima. Colors indicate annual radiation amounts. Color bar indicates radiation amounts with axis pointing toward north.

Figure 4. PV and battery systems, and EV additional costs. Left y-axis is for PV system cost and right y-axis is for battery system and EV additional costs.

Figure 5. Available EV battery capacity for V2H in the system when EVs drive three hours a day. Blue line in the bottom panel shows minimum battery capacity compared to average EV battery capacity (red line) during the daytime (7am – 7pm) available to V2H with number of EVs in the system.

Figure 6. System costs, PV and battery capacity, and NPV for Shinchi from 2020 to 2040. Individual (indiv.) analyses indicate 50 households are individually analyzed and average results are shown. Aggregated (agg.) analyses indicate 50 houses are analyzed as an aggregated system. The results of the aggregated analyses are divided by 50 to be comparable to the individual case.

Figure 7. Various indices for various technology combinations with/without FIT for Shinchi with aggregated and individual analyses.

Figure 8.  Various energy and economic indices for Kyoto district for aggregated analyses.

Figure 9. Comparison of Shinchi (residential) and Kyoto (commercial) districts for the application of technology combinations of "PV (+ battery)" and "PV + EV". Average values with sales and without sales are shown. The aggregated demand data for the districts were used.

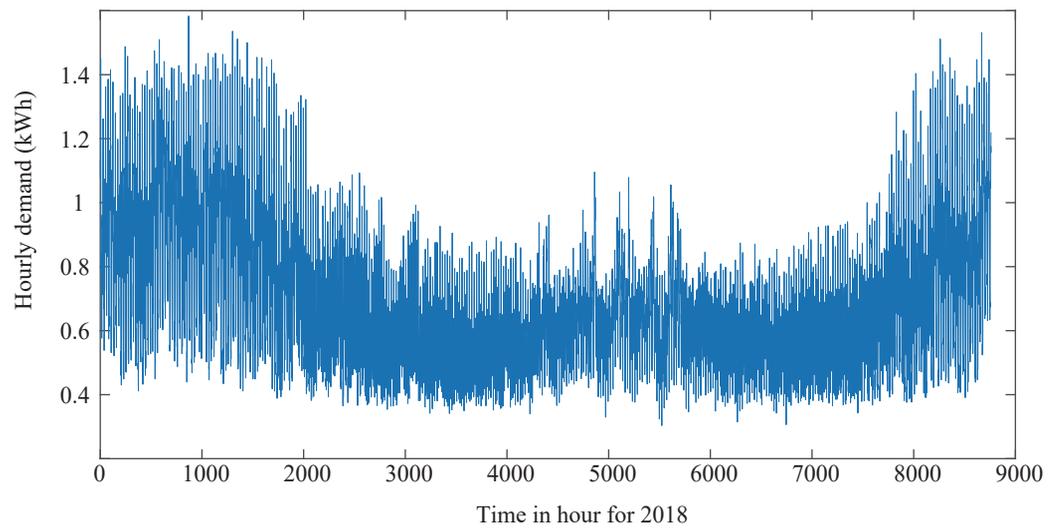

Figure 1. Average hourly demand data for 50 houses in Shinchi, Fukushima for 2018.

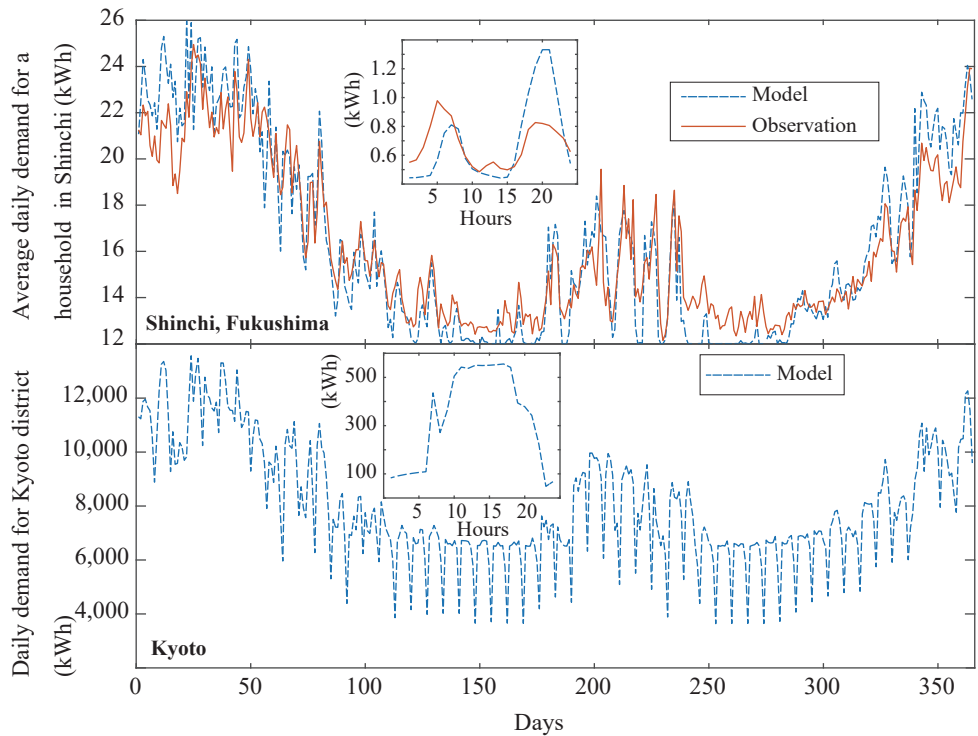

Figure 2. Daily electricity demand in Shinchi, Fukushima and Kyoto for 2018. Top panel is observed and modelled average daily demand in Shinchi from household. Bottom panel is modelled daily demand for the Kyoto district. Small windows show average hourly demands in a day for the year.

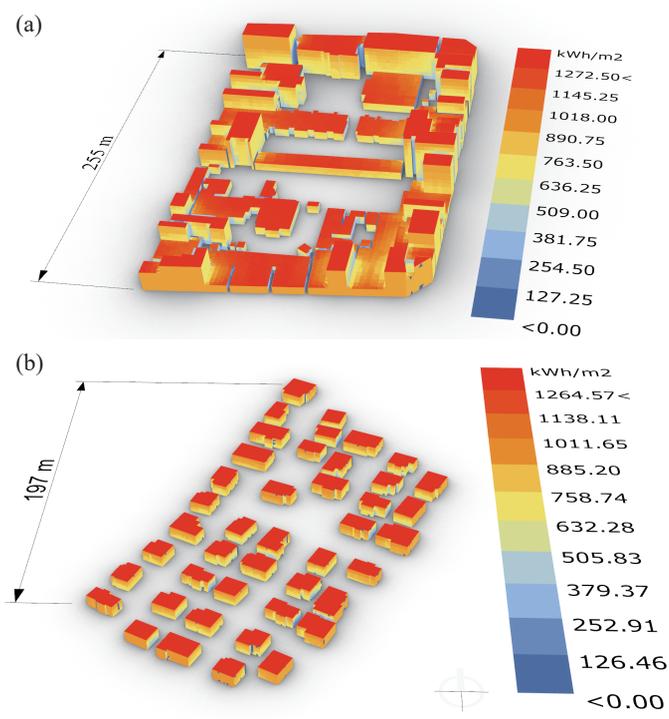

Figure 3. PV potentials (kWh/m2) on building surface areas. (a) central district of Kyoto City, (b) residential houses in Shinchi, Fukushima. Colors indicate annual radiation amounts. Color bar indicates radiation amounts with axis pointing toward north.

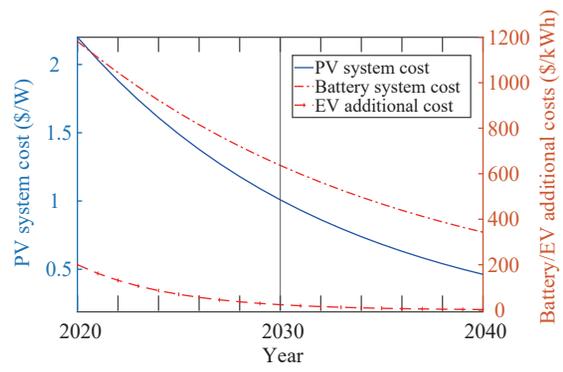

Figure 4. PV and battery systems, and EV additional costs. Left y-axis is for PV system cost and right y-axis is for battery system and EV additional costs.

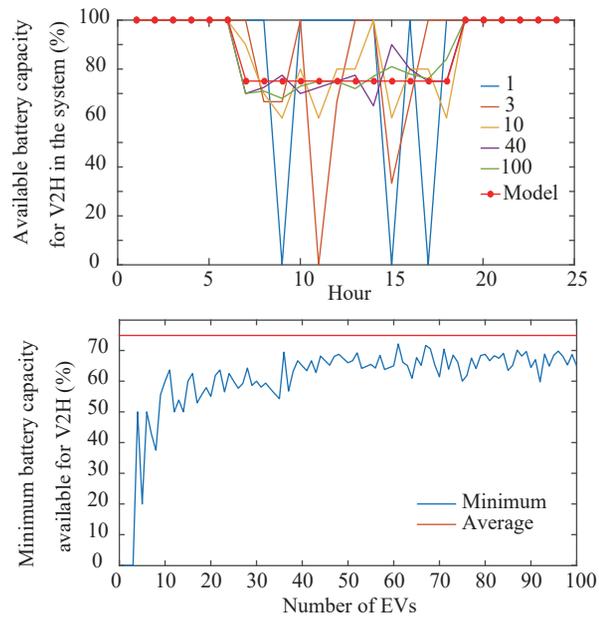

Figure 5. Available EV battery capacity for V2H in the system when EVs drive three hours a day. Blue line in the bottom panel shows minimum battery capacity compared to average EV battery capacity (red line) during the daytime (7am – 7pm) available to V2H with number of EVs in the system.

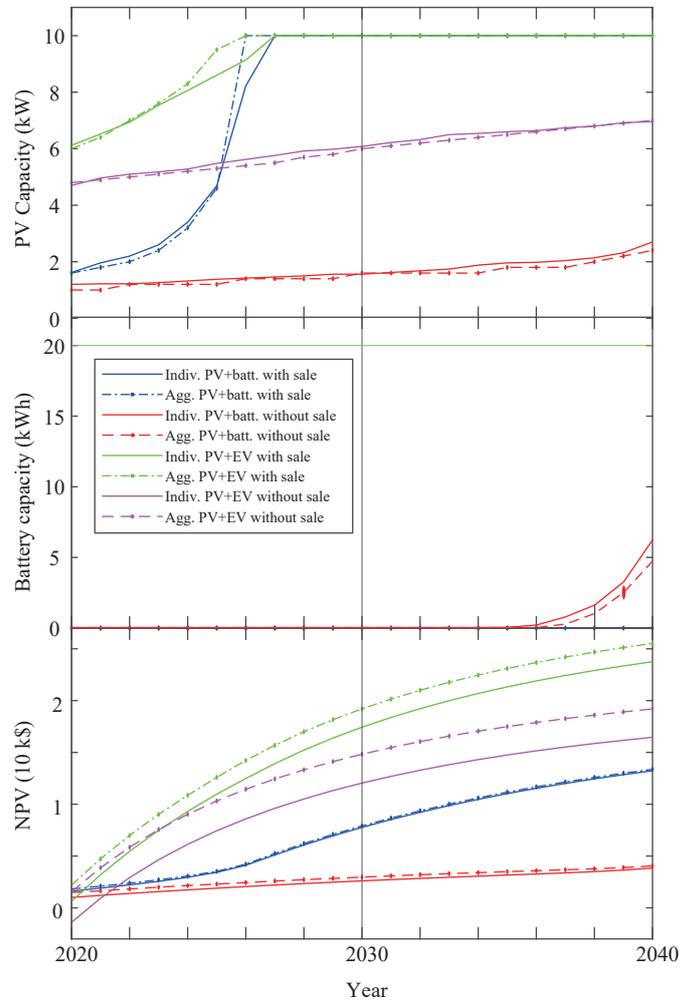

Figure 6. System costs, PV and battery capacity, and NPV for Shinchi from 2020 to 2040. Individual (indiv.) analyses indicate 50 households are individually analyzed and average results are shown. Aggregated (agg.) analyses indicate 50 houses are analyzed as an aggregated system. The results of the aggregated analyses are divided by 50 to be comparable to the individual case.

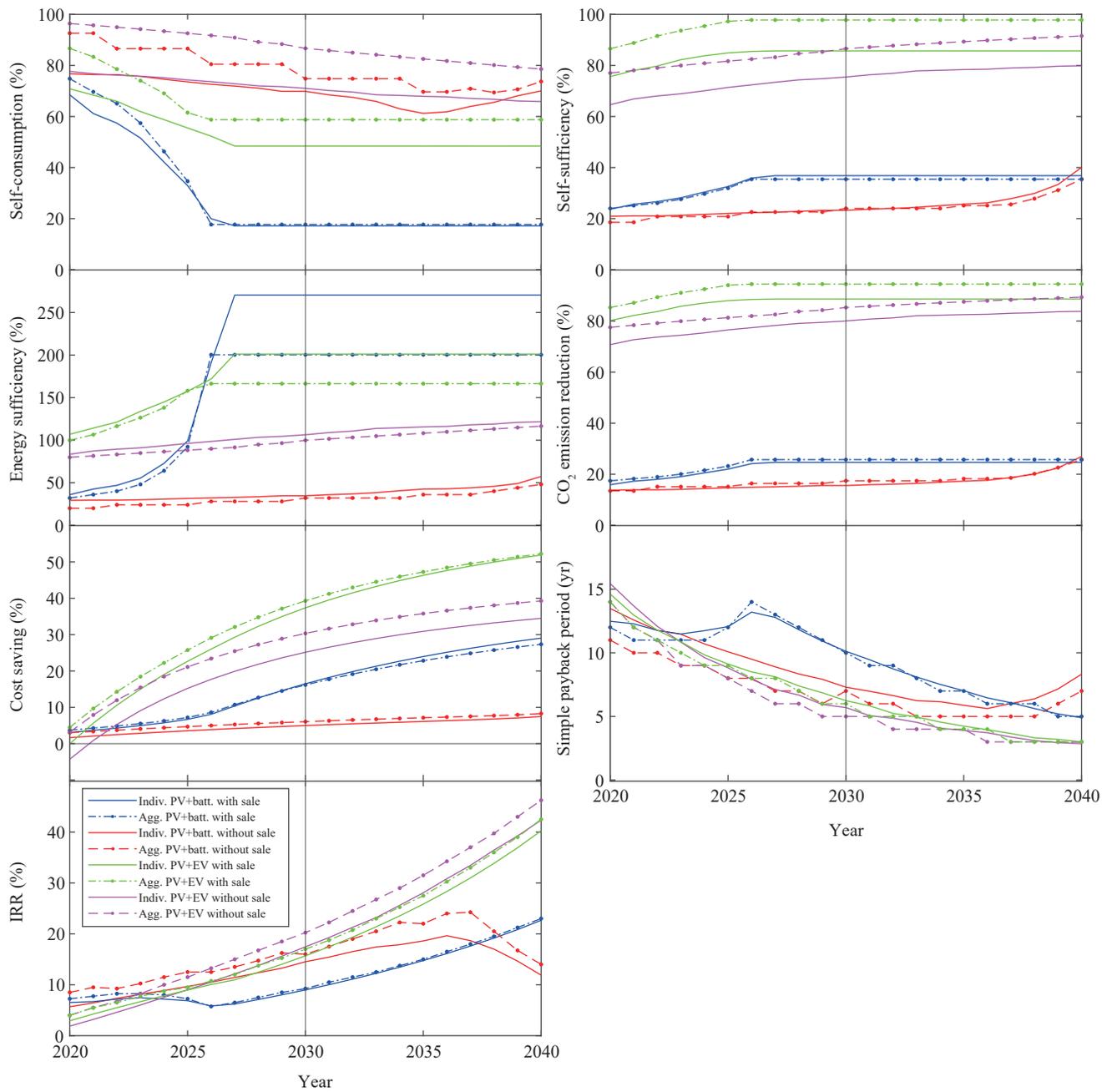

Figure 7. Various indices for various technology combinations with/without FIT for Shinchi with aggregated and individual analyses.

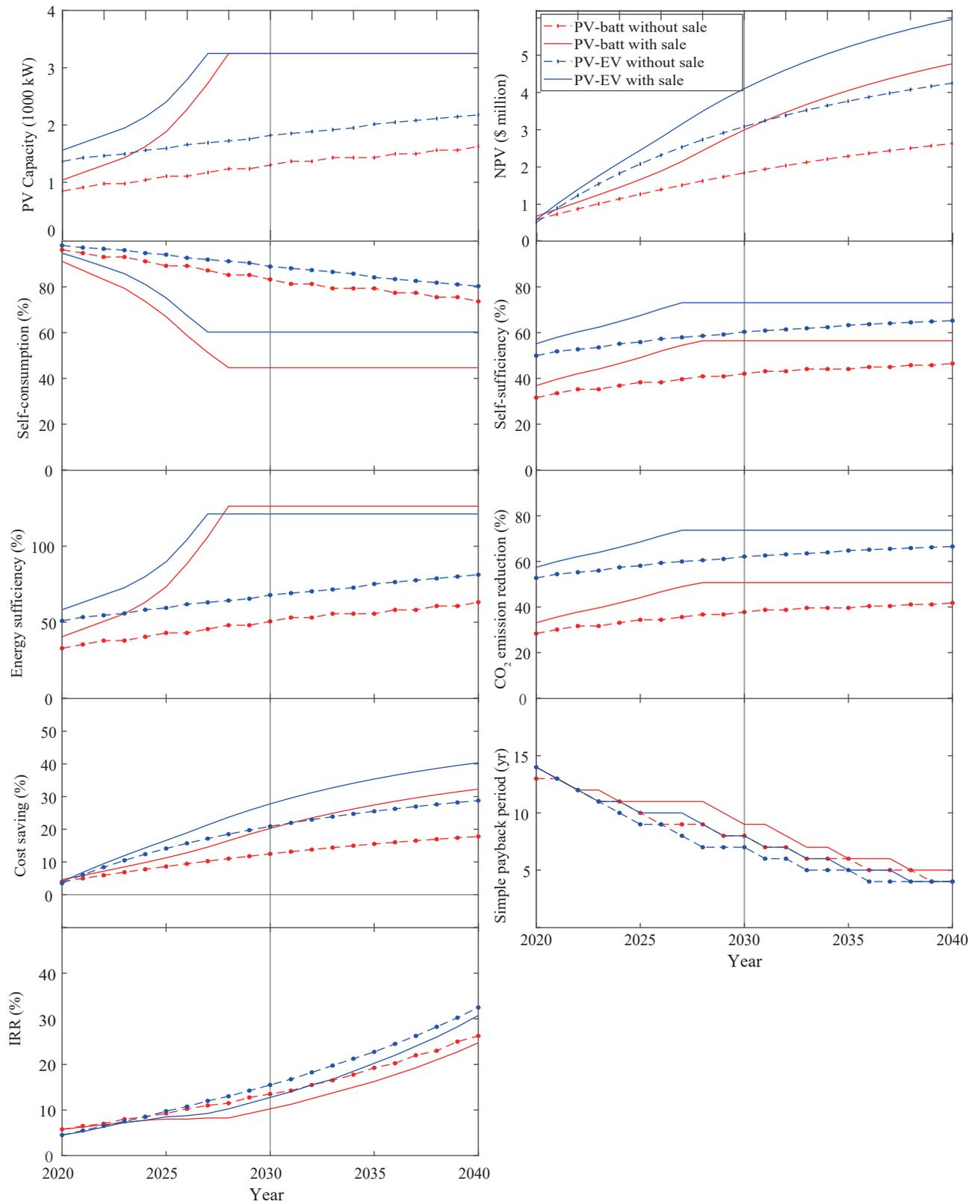

Figure 8. Various energy and economic indices for Kyoto district for aggregated analyses.

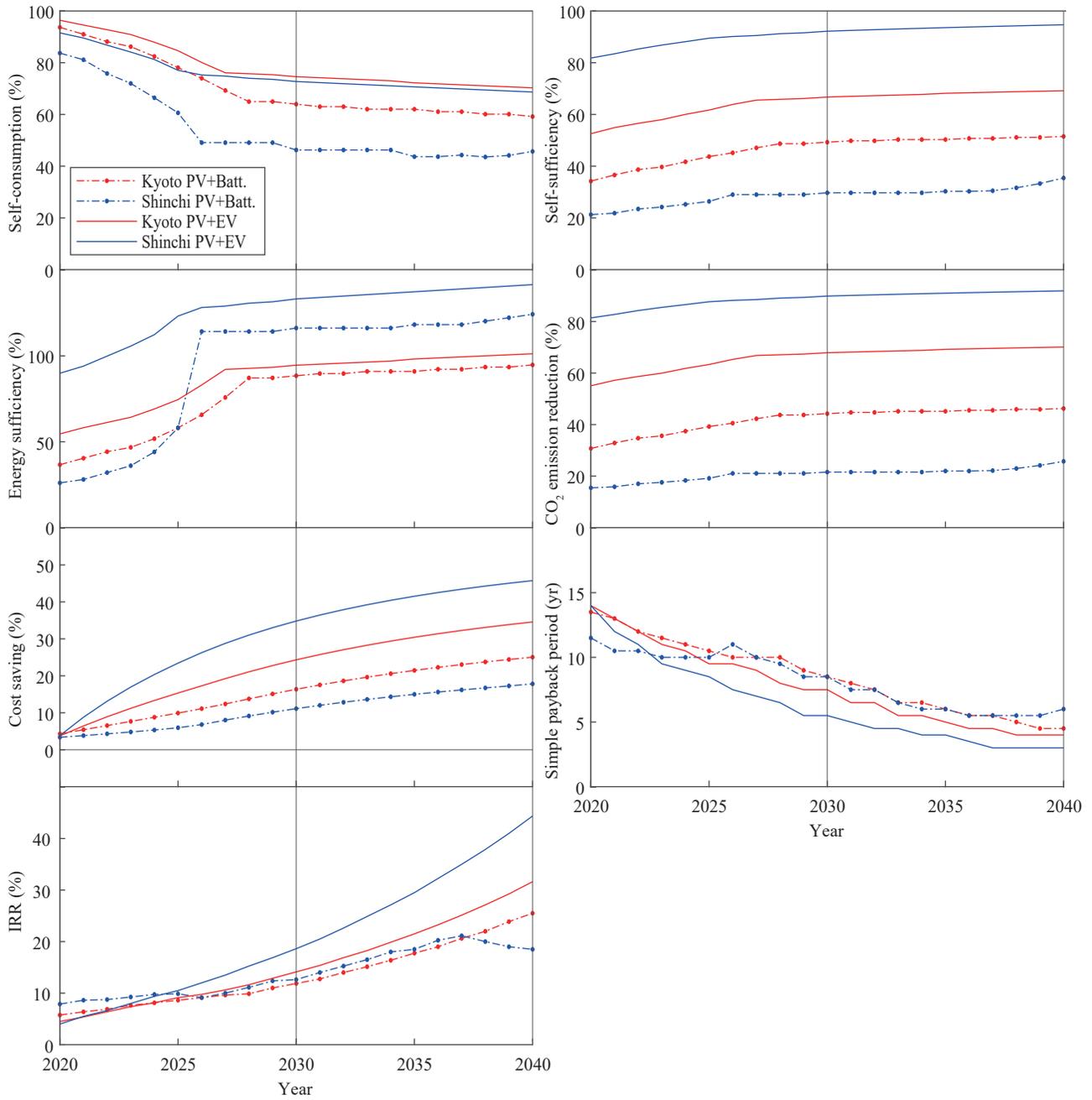

Figure 9. Comparison of Shinchi (residential) and Kyoto (commercial) districts for the application of technology combinations of "PV (+ battery)" and "PV + EV". Average values with sales and without sales are shown. The aggregated demand data for the districts were used.